\documentclass[a4paper,fleqn,usenatbib]{mnras}

\usepackage[T1]{fontenc}
\usepackage{ae,aecompl}

\usepackage{graphicx}	
\usepackage{amsmath}	
\usepackage{amssymb}	

\title[Accretion-propeller transition of neutron stars ]{THE INNER DISK RADIUS IN THE PROPELLER PHASE AND ACCRETION-PROPELLER TRANSITION OF  NEUTRON STARS}

\author[\"{U}. Ertan]{
\"{U}nal  Ertan,$^{1}$\thanks{E-mail: unal@sabaciuniv.edu}
\\
$^{1}$Sabanc\i\ University, 34956, Orhanl\i\, Tuzla, \.Istanbul,
Turkey\\
}

\date{Accepted XXX. Received YYY; in original form ZZZ}

\pubyear{2016}

\begin{document}
\label{firstpage}
\pagerange{\pageref{firstpage}--\pageref{lastpage}}
\maketitle

\def\be{\begin{equation}}
\def\ee{\end{equation}}
\def\ba{\begin{eqnarray}}
\def\ea{\end{eqnarray}}
\def\m{\mathrm}
\def\d{\partial}
\def\R{\right}
\def\L{\left}
\def\a{\alpha}
\def\acold{\alpha_\mathrm{cold}}
\def\ahot{\alpha_\mathrm{hot}}
\def\Mdotstar{\dot{M}_\ast}
\def\Omegastar{\Omega_\ast}
\def\Omegadot{\dot{\Omega}}

\def\OmegaK{\Omega_{\mathrm{K}}}
\def\Mdotin{\dot{M}_{\mathrm{in}}}
\def\Mdots{\dot{M}_{\mathrm{s}}}

\def\Mdotcrit{\dot{M}_{\mathrm{crit}}}
\def\Mdotout{\dot{M}_{\mathrm{out}}}

\def\Mdot{\dot{M}}
\def\Edot{\dot{E}}
\def\Pdot{\dot{P}}
\def\nudot{\dot{\nu}}
\def\Msun{M_{\odot}}

\def\Lin{L_{\mathrm{in}}}
\def\Lcool{L_{\mathrm{cool}}}
\def\Rin{R_{\mathrm{in}}}
\def\rin{r_{\mathrm{in}}}
\def\rlc{r_{\mathrm{LC}}}
\def\rout{r_{\mathrm{out}}}
\def\rco{r_{\mathrm{co}}}
\def\re{r_{\mathrm{e}}}
\def\Ldisk{L_{\mathrm{disk}}}
\def\Lx{L_{\mathrm{x}}}
\def\Ld{L_{\mathrm{d}}}

\def\Md{M_{\mathrm{d}}} 
\def\NH{N_{\mathrm{H}}}
\def\dEb{\delta E_{\mathrm{burst}}}
\def\dEx{\delta E_{\mathrm{x}}}
\def\Bstar{B_\ast}\def\uff{\upsilon_{\mathrm{ff}}}
\def\Bb{\beta_{\mathrm{b}}}
\def\Be{B_{\mathrm{e}}}
\def\Bp{B_{\mathrm{p}}}
\def\Bz{B_{\mathrm{z}}}
\def\Bfi{B_{\mathrm{|phi}}}
\def\BA{B_{\mathrm{A}}}
\def\tint{t_{\mathrm{int}}}
\def\tdiff{t_{\mathrm{diff}}}
\def\r_m{r_{\mathrm{m}}}

\def\rA{r_{\mathrm{A}}}
\def\BA{B_{\mathrm{A}}}
\def\rS{r_{\mathrm{S}}}
\def\rp{r_{\mathrm{p}}}
\def\Tp{T_{\mathrm{p}}}
\def\dMin{\delta M_{\mathrm{in}}}
\def\Rc{\R_{\mathrm{c}}}
\def\Teff{T_{\mathrm{eff}}}
\def\uff{\upsilon_{\mathrm{ff}}}
\def\Tirr{T_{\mathrm{irr}}}
\def\Firr{F_{\mathrm{irr}}}
\def\Tcrit{T_{\mathrm{crit}}}
\def\P0min{P_{0,{\mathrm{min}}}}
\def\Av{A_{\mathrm{V}}}
\def\ah{\alpha_{\mathrm{hot}}}
\def\ac{\alpha_{\mathrm{cold}}}
\def\tc{\tau_{\mathrm{c}}}
\def\p{\propto}
\def\m{\mathrm}
\def\fast{\omega_{\ast}}
\def\Uff{\upsilon_{\mathrm{ff}}}
\def\Ufi{\upsilon_{\fi}}
\def\Ur{\upsilon_{\mathrm{r}}}
\def\UK{\upsilon_{\mathrm{K}}}
\def\Uesc{\upsilon_{\mathrm{esc}}}
\def\Uout{\upsilon_{\mathrm{out}}}
\def\Uphi{\upsilon_{\phi}}
\def\Udiff{\upsilon_{\mathrm{diff}}}
\def\Ure{\upsilon_{\mathrm{r,e}}}
\def\U{\upsilon}
\def\UB{\upsilon_{\mathrm{B}}}
\def\tauB{\tau_{\mathrm{B}}}
\def\hA{h_{\mathrm{A}}}
\def\he{h_{\mathrm{e}}}
\def\cs{c_{\mathrm{s}}}
\def\cse{c_{\mathrm{s,e}}}
\def\hin{h_{\mathrm{in}}}
\def\rhop{\rho^{\prime}}
\def\rhod{\rho_\mathrm{d}}
\def\rhos{\rho_\mathrm{s}}
\def\rhodp{\rho_\mathrm{d}^{\prime}}
\def\rhoe{\rho_\mathrm{e}}
\def\rhoout{\rho_\mathrm{out}}
\def\Alfven{Alfv$\acute{\mathrm{e}}$n~}
\def\418{SGR 0418+5729}
\def\142{AXP 0142+61}
\def\Caliskan{\c{C}al{\i}\c{s}kan~}
\def\ql{\textquotedblleft}
\def\qr{\textquotedblright~}

\begin{abstract}
We have investigated the critical conditions required for a steady propeller effect for magnetized neutron stars with optically thick,  geometrically thin accretion disks.   We have shown through simple analytical calculations that a steady-state propeller mechanism cannot be sustained at an inner disk radius where the viscous and magnetic stresses are balanced. The radius calculated by equating these stresses is usually found to be close to the conventional \Alfven radius for spherical accretion, $\rA$. Our results show that: (1) a steady propeller phase can be established with a maximum inner disk radius that is at least $\sim 15 $ times smaller than $\rA$ depending on the mass-flow rate of the disk, rotational period and strength of the magnetic dipole field of the star, (2) the critical accretion rate corresponding to the accretion-propeller transition is orders of magnitude lower than the rate estimated by equating $\rA$ to the co-rotation radius. Our results are consistent with the properties of the transitional millisecond pulsars which  show transitions between the accretion powered X-ray pulsar and the rotational powered radio pulsar states.

\end{abstract}




\section{Introduction}

Rotational properties of a magnetized neutron star evolving with an accretion disk  are governed by the interaction between the disk and the magnetic dipole field.  
The star could spin-up or spin-down  depending on the mass-flow rate, $\Mdotin$, of the disk, strength of the magnetic dipole field, $B$, on the surface of the star and the details of the disk-magnetosphere interaction. The inner disk radius, $\rin$,  is  usually estimated by equating the viscous and the magnetic stresses and found to be close to the conventional \Alfven radius, $\rA$, which is calculated for spherical accretion by equating the ram pressure of matter to the magnetic pressure  (Lamb et  al. 1973, Davidson \& Ostriker 1973).  When $\rin$ is less than the co-rotation radius, $\rco$, at which the field lines co-rotate with the disk matter, the system is in the spin-up phase, and  the mass from the inner-disk can flow onto the star along the field lines. When $\rin$ is greater than $\rco$, the star is in the spin-down state, which is also known as the \lq propeller\rq~ phase (Illarionov \& Sunyaev 1975), since in this regime, it is usually assumed that the inflowing disk matter is expelled from the system by the dipole field lines rotating with speeds greater than the local escape speed. 

Observations of three recently discovered transitional millisecond pulsars (MSPs) which show transitions between the accretion powered X-ray pulsar and the rotational powered radio pulsar states provide an excellent laboratory to study the physics of the disk-magnetosphere interaction (see e.g. Linares  2014 for a recent review). X-ray pulsations from the transitional MSPs XSS J12270$-$4859 (Papitto et al. 2015) and PSR J1023+0038 (Archibald et al. 2015) were detected during sub-luminous  accretion states with  X-ray luminosities of  a few $10^{33} - 10^{34}$ erg s$^{-1}$ which correspond to the accretion rates of  $\sim 2 - 5\times10^{13}$ g s$^{-1}$.  The most plausible origin of these X-ray pulses seems to be  accretion onto the poles of the star (Papitto et al. 2015, Archibald et al. 2015). For $B = 2\times 10^{8}$ G, a typical field estimated for the millisecond pulsars with periods of a few ms (Lyne \& Graham-Smith 2007; see also Section 2.3),   $\rA $ is inferred to be greater than even the light cylinder radius, $\rlc = c/ \Omegastar$ where $c$ is the speed of light, and $\Omegastar$ is the angular frequency of the neutron star.  This is in sharp contrast to the expected condition,  $\rin \simeq \rco$, for accretion onto the star.  A large fraction of the inflowing disk matter might be thrown out of the system, while the observed X-ray luminosity is powered by the remaining  few per cent of the disk mass-flow accreted onto the neutron star. In this case, since $\rin$ could come close to $\rco$ accretion is possible, but the X-ray luminosity of the disk would be much higher than the observed luminosity. Is it possible to overcome this difficulty by invoking radiatively inefficient flow which could prevail when the inner disk becomes optically thin (Papitto et al 2015). This does not seem to be likely, because  the disk  becomes optically thin only when $\Mdotin$ decreases below $\sim 10^{11}$ g s$^{-1}$ (Done et al. 2007). The inner disk is expected to be optically thick at $\Mdotin \simeq 5 \times 10^{13}$ g s$^{-1}$ as inferred from the observed X-ray luminosity assuming no outflow. With a much higher $\Mdotin$ needed for $\rin \simeq \rco$, the inner disk would have an even higher optical thickness. So, one cannot invoke an optically thin disk to avoid the problem of unobserved disk luminosities if $\rin \simeq \rco$. The inferred $\rin \simeq \rA$ in transitional MSPs is in fact found to be even greater than $\rlc$. So,  how can these transitional MSPs accrete matter from the disk?

In the spin-down phase, if the inflowing disk matter cannot be repelled from the system efficiently, the resultant pile up at the inner disk increases the density and the temperature in the disk,  $\rin$ propagates inwards, and eventually reaches the co-rotation radius. Resultant onset of accretion could evacuate the inner disk, then $\rin $ moves outwards stopping the accretion. Subsequent refilling of the inner disk resuming the accretion could lead to a cyclic evolution of the inner disk even when the mass transfer from the outer disk is constant. This idea was first proposed by  Sunyaev \& Shakura (1977) and  Spruit \& Taam (1993), and recently developed through a series of papers by D'Angelo \& Spruit (2010; 2011; 2012).  In these works, it is assumed that the propeller mechanism is not efficient even for very low disk mass-flow rates. With this assumption, numerical simulations show that, for a large range of initial conditions including those with very low disk mass-flow rates, sources evolve in the long-term into the so-called trapped states, with $\rin \simeq \rco$. These states could be persistent  or the inner disk could go through cyclic accretion episodes on timescales between the dynamical and viscous timescales depending on the average mass-flow rate from the outer to the inner disk (D'Angelo \& Spruit 2012). 

The accretion and rotation powered states of the transitional MSPs do not seem to be the results of the cyclic processes described by D'Angelo \& Spruit (2012). The properties of these sources imply that there is another mechanism that impedes accretion below a critical disk mass-flow rate. 
The radio pulses are always observed below a certain X-ray luminosity less than a few $10^{32}$ erg s$^{-1}$, while $\Lx > 10^{33}$ erg s$^{-1}$ when the X-ray pulsations are observed in the sub-luminous X-ray states.  The sources remain in one or the other state on timescales (months) much longer than the dynamical or viscous timescales of the inner disk. This might indicate that the propeller mechanism is at work below a certain accretion rate. Flat radio spectra observed in the sub-luminous X-ray states could be a sign of the mass outflows (Papitto et al. 2015, Linares et al. 2014). The radio and X-ray pulses have not been observed in the same epoch, or at different epochs with the same X-ray luminosity. 
This strongly indicates that the radio pulses are switched on when the accretion is impeded, while the X-ray pulses are produced as a result of mass-flow onto the star, which switches off the pulsed radio emission.

In the disk-magnetosphere interaction model employed by D'Angelo \& Spruit (2010; 2011; 2012), the magnetosphere is formed by the closed field lines that extend up to a radius, $\r_m$.  The interaction between the disk and the dipole field takes place inside this boundary region between $\r_m$ and $\r_m + \Delta r$, while the disk is decoupled from the field beyond the interaction region. Inside the boundary, the closed field lines cannot slip through the disk, since the magnetic diffusion timescale, which is comparable to the viscous timescale (Fromang \& Stone 2009), is orders of magnitude longer than the disk-field interaction timescale. The field lines interacting with the inner disk inflate and open up. This disk-field interaction model with a narrow boundary was
proposed by Lovelace, Romanova \& Bisnovatyi-Kogan (1995), and supported by the numerical simulations (Hayashi, Shibata, \& Matsumoto  1996, Goodson, Winglee, \& Boehm 1997, Miller \& Stone 1997). We will examine the propeller effect in the context of this disk-magnetosphere interaction model which was developed further to study the disk-field interaction and the outflows in the propeller phase (Lovelace et al. 1999, Ustyugova et al. 2006). We will use the term \lq propeller state\rq  ~to indicate the phase in which the accretion onto the star is not allowed and the disk matter  arriving at the disk-magnetosphere boundary is thrown out of the system with speeds greater than the escape speed.  

In the present work, we investigate the physical conditions required for a steady or long-lasting propeller phase. In Section 2, we derive the maximum inner disk radius in a propeller regime and the critical accretion rate for the accretion-propeller transition. We discuss our results with comparisons to the characteristic properties of the transitional MSPs, and summarise our conclusions in Section 3.               

\section{Propeller Phase}

\subsection{Critical condition for the propeller mechanism}

In the case of spherical  mass-flow onto a neutron star, the matter falls with the free-fall speed, $\Uff$. The spherical mass-flow rate can be written as 
\be
\Mdots =4 \pi r^2 \rho_s(r) ~\Uff(r)          
\label{1}
\ee
where $\rho_s(r)$ is the density of the matter at radial distance $r$. Magnetic dipole field of the star can stop the mass-flow at the radius where the pressure of the falling matter becomes equal to the magnetic pressure 
\be
\frac{B^2}{8 \pi} = 
\rhos(r) ~\Uff^2 
\label{2}
\ee  
where $B = \mu / r^3$, and $\mu$ is the magnetic dipole moment of the star. equation (\ref{2}) is satisfied at the \Alfven radius
\be
\rA \simeq (G M)^{-1/7}~\mu^{4/7}~ \Mdots^{-2/7}         
\label{3}
\ee
(Lamb et al. 1973, Davidson \& Ostriker 1973) where $G$ is the gravitational constant, and $M$ is the mass of the neutron star. The inner disk radius for a geometrically thin accretion disk is is usually assumed to be the radius at which the viscous and the magnetic stresses are balanced, and calculations give inner disk radii very close to nominal \Alfven radius $\rA$ given by 
equation (\ref{3}) within a factor of $\sim 2$ (Spruit and Taam 1993). 

In the present work, we adopt the basic principles of the disk-magnetosphere interaction model proposed by Lovelace et al. (1995), and developed later by Lovelace et al. (1999) and Ustyugova et al. (2006) to study the propeller regime of magnetized stars in more detail. In this model, the radius of the axisymmetric magnetosphere, $\r_m$, is defined as the radius of the region of closed field lines. Inside $\r_m$, all matter and the field lines rotate together with the same angular velocity. The disk interacts with the large-scale field mainly in a narrow boundary region from $\r_m$ to $\r_m + \Delta r$. In the boundary region,  azimuthal  component of the magnetic field, $B_\phi$, is generated through interaction with the inner disk, and is responsible for the torque acting on the disk. Since the diffusion timescale of the field lines are much longer than the interaction timescale of the disk matter and the field, $B_\phi$ grows to a maximum, with $B_\phi \approx \Bz$, and subsequently the field lines inflate and open up (Aly 1985, Lovelace et al. 1995, Hayashi et al. 1996, Miller \& Stone 1997, Uzdensky et al. 2002, Uzdensky 2004). The matter accelerated by the magnetic torque  leaves the system along the open field lines. Reconnection of the field lines passing through the boundary layer leads to a new cycle of interaction, inflation, opening and reconnection. The outflow of the matter along the field lines is a complicated process and not well understood (see e.g. Uzdensky 2004 for a review). Here, we will simply assume that the acceleration of matter, within the interaction timescale, to the speed of the field lines rotating faster than the escape speed inside the boundary region is the minimum critical condition for an efficient propeller effect.

The timescale of interaction  between the field and the disk matter $\tint \simeq \Omegastar - \OmegaK (\rin)|^{-1}$ where $\Omegastar$ is the angular frequency of the neutron star, and    $\OmegaK (\rin)$ is the Keplerian angular speed at $r = \rin$. If the mass inflow rate into the boundary, $\Mdotin$, is constant and the  reconnections occur sufficiently rapidly leading to an average mass-outflow with a rate $\Mdotout = \Mdotin$, the system could be well approximated by a steady-state model, neglecting the small-scale rapid variations in the boundary. Indeed, the simulations indicate that the inflation-reconnection events occur on the dynamical timescale at the inner disk, $\sim 2 \pi /  \OmegaK(\r_m)$   (Lovelace et al. 1999, Ustyugova et al. 2006). In the propeller phase, when the outflowing matter  does not return back to the disk, and the outflows from outside the boundary are negligible,   the disk outside $\rin = \r_m + \Delta r$ is expected to have the properties of a standard thin disk. For a narrow $\Delta r$ that is a small fraction of $\r_m$, the inner radius of the unperturbed disk $\rin = \r_m$.  

There are a few important points that should be taken into account in the calculations of the the inner disk radius in the propeller phase to remain self-consistent with the assumed nature and timescales of the mass outflows. The matter inside the boundary is accelerated from the Kepler speed to the speed of the field lines moving with speeds greater than the escape speed on the interaction timescale, that is, the speeds do not increase gradually from $\UK$ at  $\rin$ to the speed of the field lines at $\r_m$ along many orbits. Instead the matter should be accelerated rapidly, because the interaction timescale corresponds to a small fraction of the local Keplerian period. Depending on the efficiency of reconnections of the field lines, a given piece of matter couples to the reconnected field lines at a point inside the boundary, accelerated to high speeds in the azimuthal direction,  and expelled form the system along the open field lines. As we show in Section 2.2, this radius cannot be obtained from the balance between the magnetic and the viscous stresses. Instead, the  magnetic stresses inside the boundary must satisfy the condition imposed by the interaction timescale. There is a certain amount of angular momentum that should be transferred to the matter to increase its speed above the critical level before the currently interacting closed field lines open up. The minimum magnetic torque required for this angular momentum transfer depends on how fast this should be done, that is, on the interaction timescale.   

\subsection{Escape radius}

To estimate the maximum inner disk radius in the propeller phase, we take  $B_\phi \approx \Bz$, and  $B_\phi \Bz \approx B^2 = \mu^2 / r^6$  which gives the maximum attainable magnetic torque inside the boundary. With this assumption, magnetic torque per unit radial distance of the disk becomes $B^2 r^2$ (Ghosh \& Lamb 1979). Cross sectional area of the disk perpendicular to the radial direction is $2 \pi r~ 2h$ where $H = 2h$ is the full thickness of the disk. The magnetic torque per unit volume of the disk can be written as  $\tauB = B^2  r /(4 \pi h)$ (e.g. Rappaport et al. 2004). We neglect any heating effect by the propeller mechanism inside the boundary. In the steady state, mass-flow rate of the geometrically thin disk at a radial distance $r$ can be written as
\be
\Mdot = 2 \pi r~  \Sigma(r) ~\Ur =  2 \pi r~ 2h~ \rho ~\Ur        
\label{4}
\ee
where $\Ur = |\Ur| $ is the radial velocity of the mass-flow, $\Sigma \simeq \rho H$ is the surface density, and $\rho$ is the mid-plane density of the disk.  Hydrostatic equilibrium in the direction perpendicular to the disk mid-plane gives $h \simeq \cs /\OmegaK$ where $\cs$ is the local sound speed and $\UK(r) = r~ \OmegaK(r)$ is the Kepler speed of the disk at radius $r$. Using the $\a$ prescription of the kinematic viscosity, $\nu = \a ~\cs ~h $ ~(Shakura \& Sunyaev 1973), we  have   
\be
\Ur ~\simeq ~\frac{\nu}{r} ~=~ \a~\cs~ \frac{h}{r} ~=~ \a~\frac{h^2}{r^2} ~\UK              
\label{5}
\ee 
which shows that $\UK \sim 10^5 ~\Ur$ with $h/r \sim 10^{-2}$, and $\a \sim 0.1$. This shows how slowly the matter flows in the radial direction relative to the Keplerian orbital speed.  

In the boundary region, the timescale of interaction between the inner disk matter and the field lines $\tint \simeq (\Omegastar - \OmegaK (\rin))^{-1} = (\omega - 1)^{-1} ~\OmegaK(\rin)^{-1}$  where $\omega = \Omegastar / \OmegaK (\rin)$ is the fastness parameter. The escape speed at a given radius $\Uesc = \sqrt{2} \UK$.  To expel the matter from the system,  the speed of the field lines should be greater than $\Uesc$. This is satisfied outside the radius where  $\omega = \Omegastar / \OmegaK (\rin)  =    \sqrt{2}$. At  $r = \rco$ the field lines and the matter move with the same angular speed. Since  $\OmegaK \p r^{-3/2}$ ,   the propeller mechanism could be sustained with an inner disk radius greater than  $(\sqrt{2})^{2/3} \rco \simeq 1.26~ \rco$. In the boundary region, the magnetic torque acting on the matter should accelerate the matter to the speed of the field lines within the interaction timescale, $\tint$.This means that the magnetic stresses should  accelerate the matter with the rate 
$\Omegadot \simeq \Delta \Omega / \tint$ where $\Delta \Omega = \Omegastar - \OmegaK (\rin)$, which requires an angular momentum transfer to the unit volume of  the matter with the rate  
\be
\rho r^2 ~\Omegadot  \simeq \rho r^2 ~\frac{\Delta \Omega}{(\Omegastar - \OmegaK)^{-1}}~= ~(\omega - 1)^2~ \rho r^2 ~\OmegaK (\rin)^2. 
\label{6}
\ee 
Then, the magnetic torque per unit volume of matter, $\tauB$, should satisfy the critical condition 
\be
\tauB = \frac{B^2  r}{4 \pi h} \simeq ~(\omega - 1)^2~ \rho r^2 ~{\OmegaK (\rin)}^2
\label{7}
\ee 
for a steady propeller effect. Apart from the $r/h$ term on the left hand side, equation (\ref{7}) looks similar to equation (\ref{2}) which gives the conventional \Alfven radius  with the free-fall speed $\Uff \sim r \OmegaK$. Nevertheless,  for a given accretion rate at a radius $r$, the mass density in the disk is $(r /h) (\UK/ \Ur) \simeq  \a^{-1} (r/h)^3 \sim 10^7$ times greater than the mass density in the spherical accretion. This means that the $B^2$ value at the conventional \Alfven radius is about 5 orders of magnitude lower than that required for the propeller effect. To put it in other words, the inner disk radius  in a steady propeller phase is likely to be much smaller than the \Alfven radius given by equation (\ref{2}).  

Using  equations (\ref{3} - \ref{5}) and (\ref{7}), we obtain the maximum inner disk radius in terms of the conventional \Alfven radius   
\be
\rin \simeq \left[~(\omega - 1)^{-2}~ \left( \a ~\frac{h^2}{r^2}\right)  \right]^{2/7} ~\rA.
\label{8}
\ee  
Using the notation $\a_{-1} = \a /0.1$, and $(h/r)_{-2} = (h/r)/ 10^{-2} $, equation (\ref{8}) can be rewritten as
\be
\rin \simeq 3.8 \times 10^{-2} ~\a_{-1}^{2/7} ~(\omega - 1)^{-4/7}~ \left(\frac{h}{r}\right)_{-2}^{4/7} ~\rA
\label{9}
\ee  
where $h/r$ should be calculated at $r=\rin$ for the unperturbed disk. 
Note that  $h/r \p \Mdotin^{3/20} ~r^{1/8}$ has a weak dependence on both $r$ and $\Mdotin$ ~(see e.g. Frank et al. 2002). We would like to stress that equation (\ref{9}) is valid only in the propeller phase, that is, when $\omega > \sqrt{2}$. In this regime, the $\rin / \rA$ ratio  increases with decreasing $\omega$, and reaches a critical value corresponding to propeller-accretion transition when $\omega \simeq \sqrt{2}$. , Using equation (\ref{3}), we obtain 
\be
\left(\frac{\rin}{\rA}\right)_{\mathrm{crit}} \simeq 6.3 \times 10^{-2} ~\a_{-1}^{2/7} ~ \left(\frac{h}{r}\right)_{-2}^{4/7}.
\label{10}
\ee  
For large ranges of  $\Mdotin ~(10^{12} - 10^{16}$ g s$^{-1}$) and $r ~(10^7 - 10^{10}$ cm), the $h/r$ ratio remains between $\sim 2\times 10^{-3} -  10^{-2}$. 
For  $\a = 0.1 $, $\rA$ is found to be $\sim 15 - 40$ times greater than $\rin$ during the accretion-propeller transition. 

The boundary layer is a region of continual opening and reconnections of the field lines. From this region the matter leaves the system along the open field lines (see e.g. Lovelace at al. 1999 for a detailed description of the model).  The outer boundary has a surface area $A = 2 \pi \rin 2h$. The mass per unit volume passing through this surface will rapidly gain an angular momentum $\simeq \tauB \tint$ through interaction with the field lines within $\tint$.  The disk matter enters the boundary with radial  speed $\Ur$ much smaller than its azimuthal speed $\Uphi \simeq \UK$. Inside the boundary, that matter is accelerated in the azimuthal direction to the speed $\rin \Omegastar = \omega \rin \OmegaK$ within $\tint = (\omega - 1)^{-1}  ~\OmegaK(\rin) ^{-1}$. Using equation (\ref{4}) with $\Mdotin = \Mdotout$, the total angular momentum transfer rate from the field to the outflowing matter becomes   
\be
\dot{L}  \simeq ~\tauB  ~\tint ~\Ur ~A \simeq~\Mdotin ~(\omega - 1)~ \rin^2 ~\OmegaK(\rin)
\label{11}
\ee  
when $\r_m \simeq \rin > 1.26 ~\rco$.  This could be taken as a lower limit for the torque acting on the star, since we have neglected the effects of possible outflows returning back to disk which might lead to a pile up at the inner boundary. All these effects tend to further decrease the inner disk radius in the propeller phase. The inner disk radius in a steady-state cannot be built up at radii greater than $\rin$ given by equation (\ref{9}). This is because the required angular momentum for the matter to escape should be transferred within the interaction timescale. In particular, at the \Alfven radius, this needs a time interval that is about $10^5$ interaction timescale. Initially, the matter could be stopped at $\sim \rA$, but since the matter cannot be accelerated to high speeds there,  the matter piles up at the inner disk. This extends the inner disk inwards until the critical condition for a steady propeller effect is satisfied at the inner disk radius. When the system reaches the steady-state, the disk regions outside the boundary region remain decoupled from the field lines, and thereby in an unperturbed steady thin disk state.  We have proposed that this steady state could be achieved with an inner disk radius less than around  $\rin$ given in equation (\ref{9}).

\subsection{The critical accretion rate for accretion-propeller transition}

The critical mass-flow rate of the disk, $\Mdotin$, for transition between the accretion and the propeller phases could be estimated by setting $\rin =1.26 \rco$ and $\omega = \sqrt{2}$ in equation (\ref{9}). Using also equation (\ref{3}), and $\rco = (GM)^{1/3}  ~\Omegastar^{-2/3}$, we find
\be
\Mdotcrit \simeq 4 \times 10^{12} ~g s^{-1} ~\a_{-1} ~P_{-3}^{-7/3} ~ \mu_{26}^2 \left(\frac{h}{r}\right)_{-2}^{2}
\label{12}
\ee  
where $\mu_{26}$ is the magnetic dipole moment of the star in units of $10^{26}$ G cm$^3$, and $P_{-3}$ is the spin period of the star in units of milliseconds.  The propeller phase could prevail up to a maximum mass inflow rate of the disk given by equation (\ref{12}), which corresponds to critical accretion and disk luminosities $\Lx = G M \Mdot / R $ and $\Ld = G M \Mdot / \rin$ respectively, with $\Mdot = \Mdotcrit$.

Recently, Joadand et al. (2016) showed through detailed timing analysis that   J1023+0038  slows down in the X-ray pulsar phase with a rate $\nudot = -3.0413 \times10^{-15}$ Hz s$^{-1}$) which is about $27 \%$ higher than the rate in the radio-pulsar phase ($\nudot = -2.3985\times10^{-15}$ Hz s$^{-1}$).  This small fraction of increase in the torque when the system undergoes a transition from the radio pulsar to X-ray pulsar state might indicate that the magnetic dipole torque dominates the propeller torque in the radio  pulsar phase. For a purely magnetic dipole torque, the dipole field strength on the pole of the star can be estimated through  $B\sim 6.4 \times 10^{19} \sqrt{P \Pdot}$ G.  For J1023+0038, which has very similar properties to other transient  sources, observed period and period derivative give $B = 2 \times 10^8$ G (half of this at the equator) which is a typical field also for the other millisecond pulsars with shortest periods  (see e.g. Lyne and Graham-Smith 2007).

For a MSP with $P=1.69$ ms and $B \simeq 2\times 10^8$ G on the surface of the star, the condition  $\rA = 1.26 ~\rco$,  is obtained with an accretion rate of $\sim 6 \times 10^{16}$ g s$^{-1}$ (equation (\ref{3})), which is about 4 orders of magnitude greater than the critical accretion rate indicated by  equation (\ref{12}). It is seen that $\Mdotcrit \propto P^{-3/7}$. This implies that the systems with smaller periods have higher critical accretion luminosities. That is, for a given field strength,  the systems with relatively low periods undergo accretion-propeller transition at higher X-ray luminosities. This could be the reason for that the observed transitional  MSPs to be short-period systems. The critical accretion rate given in  equation (\ref{12}) is consistent with the properties of MSPs during the transitions between the X-ray pulsar and the radio pulsar states (see Section 3 for further discussion).  
 
\section{DISCUSSION AND CONCLUSIONS}

We have investigated  the required conditions for a steady-state  propeller phase, for which all the matter arriving at the inner disk radius $\rin > 1.26 ~\rco$  can be expelled from the system with $\Uout > \Uesc$ and $\Mdotin = \Mdotout$.  We have found that the propeller state could be established with an inner disk  radius more than $\sim15$ times smaller than the  conventional \Alfven radius depending on the dipole field strength and the period of the star, and the mass-flow rate of the disk.

We derive $\rin$ with the assumption that the disk outside the boundary has the properties similar to a standard thin disk. For greater surface densities due to some weak interactions between the field lines and the disk outside the boundary region,  the inner disk would then settle down with an inner radius even smaller than the radius given by equation (\ref{9}). We take the maximum possible magnetic stresses in our calculations; similarly,  weaker magnetic stresses would give relatively small inner disk radii.   Nevertheless, considering very sharp radius dependence of the magnetic torques the actual inner disk radius, in the propeller regime, is likely to be near the maximum possible inner disk radius $\rin$ given by equation (\ref{9}).  

When the orbital kinetic energy difference of the matter with $\UK$  and $\rin \Omegastar$ is larger than the thermal energy density, Kelvin-Helmholtz instabilities become important for the gas to penetrate into the boundary region (Spruit and Taam 1993).  The diffusion timescale of the disk into the boundary region $\tdiff \sim \Delta R / \Udiff$ where  $\Udiff$  is the minimum of  $\Delta \U \sim \rin (\Omegastar - \OmegaK)$~ and $\cs$. For radii greater than $1.26 \rco $,   $\Delta \U$ is much greater than $\cs$, and $\Ur \ll \cs$ for a thin disk.
This means that the inner disk can  penetrate the boundary region without forming a pile-up in front of the boundary in the propeller phase. Furthermore, all the matter flowing into the boundary can be expelled from the boundary with the same rate as the rate of mass-flow into the boundary. This picture is consistent with our standard thin-disk assumption.

In the numerical simulations, the efficiency and the time dependence of the outflows depend on some model assumptions (see Uzdensky 2004 for a  review). In the present work, to estimate the critical condition for the propeller phase, we have adopted the basic principles of the disk-magnetosphere interaction model proposed initially by Lovelace et al. (1995), and developed further by Lovelace et al. (1999) to study the outflows in different regimes. The details of the propeller phase in the same model was studied through numerical simulations by Romanova et al. (2005) and Ustyugova et al. (2006). In these simulations, the magnetosphere is defined as the region in which all the matter and the closed field lines rotate together with the angular speed of the star while any plasma flows are along the field lines. The magnetospheric radius is taken to be equal to the radius at which $B^2 / 8 \pi \simeq  P+\rho \Uphi^2$ in the boundary where the thermal pressure, $P$, is negligible under the conditions we consider here.  The matter inside the boundary is accelerated from quasi-Keplerian speeds. The inner disk radius $\rin$, which is not known a priori, is determined from the simulations. 
The inner disk radius is always found to be close to the magnetospheric radius ($\rin \simeq  \r_m$) within a factor of  $\sim 3$ in the propeller phase even when the systems show different outflow morphologies (Ustyugova et al. 2006).  In line with these results,  the inner disk radius, $\rin$, given in equation (\ref{7}) remains close to $\r_m$  found by Lovelace et al. (1999) for all reasonable values of $\omega$, except for when the system is very close to the propeller-accretion transition ($\rin = 1.26 \rco$).   From equations (\ref{7}) and  (\ref{8}), it can be shown that $\rin  \simeq (\omega - 1)^{-4/7} (r/h)^{2/7} \r_m$. For $r/h \sim 10^2$, the $\rin / \r_m$ ratio changes from 3.7 to 1 for the values of      $(\omega - 1) $ from 1 to 10.  Nevertheless, for $\omega$ values greater than $\sim 10$, our $\rin$ decreases slightly below $\r_m$ found  by Lovelace et al. (1999).  For instance, $\omega = 20$ gives  $\rin \simeq 0.7 \r_m$.   

The best sources to test our results are the recently discovered and clearly observed transitional MSPs which show transitions between the accretion powered X-ray pulsar and the rotational powered radio pulsar states. Two of these sources, namely PSR J1023+0038 (Archibald et al. 2015) and XSS J12270$-$4859 (Papitto et al. 2015), are observed to show X-ray pulsations in the sub-luminous X-ray states with $\Lx \approx ~$a few$~ 10^{33} - 10^{34}$ erg s$^{-1}$.  Properties of these X-ray pulsations strongly indicate that they are powered by mass-flow onto the neutron star. These sources are observed in the radio pulsar phases when $\Lx \sim$ ~a few$~ 10^{32}$ erg s$^{-1}$ with no X-ray pulsations. It seems that the radio pulsar phase is switched on at a disk mass-flow rate lower than $\sim 10^{13}$ g s$^{-1}$, and above this X-ray luminosity the sources are in the accretion phase.  Since the accretion switches off the pulsed radio emission, the sources are very likely to be in the strong propeller phase when they emit radio pulses. Using the critical condition $\rin \simeq 1.26 ~\rco$ for accretion-propeller transition, we can estimate the X-ray luminosity originating from the accretion onto the star, $\Lx$, just before the transition to the propeller state and the X-ray luminosity of the disk, $\Ld$, just after the transition. The period  $P = 1.69$ ms of the transitional MSPs corresponds to  $\rco \simeq 2.4 \times 10^{6}$ cm. For a neutron star with radius $R = 10^6$ cm and mass $M = 1.4 \Msun$,  the transition luminosity $\Lx  = G M \Mdot /  R  \simeq 1\times 10^{33}$ erg s$^{-1}$ corresponds to an accretion rate $\Mdot \sim 5 \times 10^{12}$ g s$^{-1}$. Immediately after the transition to the propeller phase, the X-ray luminosity decreases to $\Ld  = G M \Mdotin / 2 \rin \sim 1.5 \times 10^{32}$ erg s$^{-1}$.  Note that these estimates are independent of the model details, except for the assumed inner disk position during the transition.  

Is $\rin$ indeed close to $1.26 ~\rco$ for  $\Mdot \sim 5 \times 10^{12}$ g s$^{-1}$? In other words, is the critical accretion rate corresponding to this $\rin$ in the models, consistent with this critical accretion rate?  In our model, we can check this from equation (\ref{12}). For $P$ = 1.69 ms  and $B \sim 10^{8} - 10^{9}$ G, the values of the critical accretion rate $\Mdotcrit$ calculated from equation (\ref{12}) are found to be consistent with the transition rate indicated by the observations.  Considering the possibilities of the interaction between the pulsar wind and the inner disk, and the magnetic heating in the disk-field interaction region, the observed X-ray luminosity in the radio pulsar phase could be taken as an upper limit to the standard disk luminosity.

 Our result for the inner disk radius is applicable in the propeller phase  ($\rco < 1.26 \rin$).  In the spin-down with accretion phase, when $\rco < \rin < 1.26 ~\rco$,  the matter cannot be thrown out with sufficiently high speeds and piles up at the inner disk. This could lead to 
repetitive switches  between the accretion and the propeller phases on dynamical and viscous timescales of the innermost disk. These transitions might be responsible for the observed rapid recurrent changes between two different X-ray modes of the transitional MSPs with luminosity ratios of $\sim 5 -7$ (see Linares 2014 for a review).  It is remarkable that $\Lx / \Ld \simeq 6$ when  the inner disk radius is $\rin \sim 1.26 ~\rco \simeq 3 \times 10^6$ cm. This phase, neglecting the propeller effect, was studied in detail by D'Angelo and Spruit (2012). In this regime, the total torque acting on the star is also expected to be rather different from that in the propeller phase. Because, the pile up at the inner disk required to satisfy $\rin = \rco$ increases the viscous stresses considerably. Furthermore, the interaction timescale of the inner disk matter and the field lines increases as $\rin$ approaches to $\rco$, which makes the transition to the spin-up phase rather complicated together with  uncertainties  in the radial width of the boundary region. We will investigate the properties of the spin-down phase with accretion and the transition to the spin-up phase in an independent work. 

In summary, we have estimated the maximum inner disk radius in the propeller phase and the critical accretion rate for the accretion-propeller transition for magnetized stars with optically thick geometrically thin standard accretion disks. We have shown that a steady propeller phase could be achieved with an inner disk radius much lower than the conventional \Alfven radius. In this phase, the inner disk radius decreases with increasing mass-flow rate of the disk. At a critical accretion rate, corresponding to $\rin \simeq 1.26 ~\rco$, the system makes a transition from the propeller to the accretion phase. This accretion rate is much lower than the rate that can balance the magnetic and the viscous stresses close to the co-rotation radius. The transition from the accretion into the propeller phase  is  likely to be accompanied by the switch-on of the pulsed radio emission, while all the mass flowing into the boundary region is being effectively thrown out along the open field lines.  Our results, that are in agreement with the properties of transitional MSPs, can also be extended to other magnetized stars accreting from geometrically thin accretion disks.


\section*{Acknowledgements}

 We acknowledge research support from
T\"{U}B{\.I}TAK (The Scientific and Technological Research Council of
Turkey) through grant 113F166 and from Sabanc\i\ University.  We thank Ali Alpar for useful comments on the manuscript.  












\bsp	
\label{lastpage}
\end{document}